\documentclass[english]{appolb}
\usepackage{graphicx}
\usepackage{epsfig}
\usepackage{babel}
\usepackage{amsmath}
\usepackage{amssymb}

\usepackage{cite}
\begin{document}
\title{
The Casimir-Polder interaction between two neutrons and possible relevance 
to tetraneutron states
\thanks{Presented by MSH at the 2nd Jagiellonian Symposium on Fundamental and Applied Subatomic Physics, June 4 - 9, 2017, Krak\'{o}w}%
}
\author{
{M. S. Hussein}
\address{Instituto Tecnol\'{o}gico de Aeron\'{a}utica, DCTA,12.228-900 
S\~{a}o Jos\'{e} dos Campos, SP, Brazil}
\address{Instituto de Estudos Avan\c{c}ados, Universidade de S\~{a}o Paulo 
C. P. 72012, 05508-970 S\~{a}o Paulo-SP, Brazil}
\address{Instituto de F\'\i sica, Universidade de S\~ao Paulo,
R. do Mat\~{a}o 1371, 05508-090, S\~ao Paulo, SP, Brazil}
\\
{J. Babb}
\address{ITAMP, Harvard-Smithsonian Center for Astrophysics, MS 14, 60 
Garden St., Cambridge, MA 02138, USA}
\\
{R. Higa}
\address{Instituto de F\'\i sica, Universidade de S\~ao Paulo,
R. do Mat\~{a}o 1371, 05508-090, S\~ao Paulo, SP, Brazil}
}
\maketitle
\begin{abstract}
We present a summary of our recent publication concerning the derivation of 
the extended Casimir-Polder (C-P) dispersive interaction between two neutrons. 
Dynamical polarizations of the neutrons, recently derived within Chiral 
Effective Theory are used for the purpose. An account of the higher 
frequency/energy behavior of these entities related to the opening of 
one-pion production channel and the excitation of the $\Delta$ resonance 
are taken into consideration in our derivation of the C-P interaction. 
The neutron-neutron system in free space is treated in details so are the 
neutron-wall and the wall-neutron-wall systems. The case of tetraneutron 
(a 4 neutron system) in a resonant state is then briefly considered. 
The 4n C-P interaction is evaluated to assess its potential relevance to 
the ongoing debate concerning the nature of the tetraneutron.
\end{abstract}


\section{Introduction} 
It is by now a given fact that hadrons are bound entities of fractionally 
charged quarks and the ensuing effects of this picture results in their 
electric and magnetic polarizabilities. The long-distance strong interaction 
between neutrons arise from exchanges of quarks and gluons in colorless 
states (bosons) driven mainly by the underlying chiral symmetry of quantum 
chromodynamics (QCD). The electromagnetic interactions between  well-separated neutrons arise 
from the Casimir-Polder  (C-P) effect related to the dipole polarizability 
\cite{casimir48,bernabeu76,babb10}. Knowing the C-P interaction involving neutrons is 
important from the practical point of view as it has relevance in ultracold 
neutron physics, and in particular in the understanding of the working of 
neutron confining bottles. In this contribution we give a summary of our 
recent publication on the subject and also comment on the tetraneutron.

\section{C-P interaction between two neutrons}
Recently, we have investigated the neutron-neutron dispersive Casimir-Polder 
interaction between two neutrons \cite{BHH17}. In that work, we assessed the 
importance of internal excitation of the neutron and pion production on the 
dynamic polarizabilities. We present in the this section our results for 
the n-n, C-P interaction.

As shown by Feinberg and Sucher~\cite{FeiSuc70}, the asymptotic 
($r\sim\infty$) long-distance electromagnetic interaction between two 
neutrons is given by the Casimir-Polder potential, 
\begin{equation}
\label{nn-infty}
V^\infty_{CP,nn}(r) =  - \frac{\hbar c}{4\pi r^7}
[23(\alpha_{n}^2 + \beta_{n}^2) -14\alpha_{n}\beta_{n} ] + \mathcal{O}(r^{-9})
=V^{*}_{CP,nn}(r) + \mathcal{O}(r^{-9}),
\end{equation}
with the notation $V^{*}_{CP}$ meaning the static limit of the nucleon 
dynamic polarizabilities. 
Bernab\'eu and Tarrach, on the other hand, derived the analogous long-range 
potential between a proton and a neutron~\cite{bernabeu76}, 
\begin{eqnarray}
\label{np-infty}
V^\infty_{CP,pn}(r) &=& \hbar c\,\alpha_0\left[ 
-\frac{\alpha_{n}}{2r^{4}}
+\frac{1}{4 \pi c M_p r^{5}}(11\alpha_n + 5\beta_n) 
+ \mathcal{O}(r^{-7}) \right]
\nonumber\\
&=&V^{*}_{CP,pn}(r)  + \mathcal{O}(r^{-7}),
\end{eqnarray}
where $M_p$ is the proton mass and $\alpha_0=e^2/4\pi\sim 1/137$ is the 
electromagnetic fine structure constant.
It exhibits the leading repulsive $r^{-5}$ term from the polarizabilities of 
the neutron induced by the charge of the proton, followed by the $r^{-7}$ 
interaction bilinear in the two nucleon polarizabilities. 

In Ref.~\cite{BHH17} we improved on the above description by considering 
the frequency dependence on the so-called dynamical dipole polarizabilities. 
At distances large enough that exchange forces can be neglected, the 
Casimir-Polder interaction between two neutrons follows 
from~\cite{babb10,FeiSuc70},
\begin{equation}
\label{CP-dip-dip}
V_{CP,ij}(r) = - \frac{\alpha_0}{\pi r^6} I_{ij}(r)
\end{equation}
where
\begin{eqnarray}
\label{eq:integ_ij}
&&I_{ij}(r) = \int_{0}^{\infty} d\omega e^{- 2\alpha_{0} \omega r} \Big\{
\big[\alpha_i(i\omega)\alpha_j(i\omega)+\beta_i(i\omega)\beta_j(i\omega)
\big]P_E(\alpha_{0} \omega r)
\nonumber\\[1mm]&&
\hspace{2.5cm}
+\big[\alpha_i(i\omega)\beta_j(i\omega)+\beta_i(i\omega)\alpha_j(i\omega)
\big]P_M(\alpha_{0} \omega r)
\Big\},
\nonumber\\[3mm]&&
P_E(x) = x^4 + 2x^3 + 5x^2 + 6x + 3 ,
\quad
P_M(x) = - ( x^4 + 2x^3 + x^2 ) ,
\end{eqnarray}
where $\alpha_i(\omega)$ and $\beta_i(\omega)$, respectively, are the dynamic electric 
and magnetic dipole polarizability of particle $i$,  and similarly for 
particle $j$. 

Theoretical and experimental studies on the nucleon polarizabilities 
have a long tradition in hadron physics, as they unravel important 
information about the internal structure of hadrons 
(for a review see~\cite{Hagelstein:2015egb}). Low-energy analyses with  photon 
energies up to the excitation of the $\Delta$ resonance were  performed 
within the effective theory of QCD in such regime, namely, chiral effective 
field theory~\cite{Hagelstein:2015egb,Hildebrandt:2003fm,Hil05,GriMcGPhi12,Lensky:2015awa}.
 The energy-dependence of the neutron dipole 
polarizabilities $\alpha_n(\omega)$, $\beta_n(\omega)$, involve long 
expressions and integrals that are far from simple. 
Given this, in~\cite{BHH17} we proposed a parametrization with a simpler 
form that takes into account the one-pion production cusp and the $\Delta$ 
resonance contribution, 
\begin{eqnarray}
\alpha_{n} (\omega) &=& 
\frac{\alpha_n(0)\,\sqrt{(M_{\pi}+a_1)(2M_{n}+a_2)}(0.2a_2)^2}
{\sqrt{(\sqrt{|M_{\pi}^2-\omega^2|}+a_1)(\sqrt{|4M_{n}^2-\omega^2|}+a_2)}
\big[|\omega|^2+(0.2a_2)^2\big]}\,
\label{eq:edip-polariz1}
\\[3mm]
\beta_{n} (\omega) &=& 
\frac{\beta_n(0)-b_1^2\omega^2+b_2^3{\rm Re}(\omega)}
{(\omega^2-\omega_{\Delta}^2)^2+|\omega^2\Gamma_{\Delta}^2|}\,,
\label{eq:mdip-polariz1}
\end{eqnarray}
with fitting parameters $a_1$, $a_2$, $b_1$, $b_2$, $\omega_{\Delta}$, and $\Gamma_{\Delta}$.
The parameter $a_1$ is formally a higher-order effect, but  is important 
to match the correct pion production threshold, which controls the low-energy 
behavior of $\alpha_n(\omega)$~\cite{Hildebrandt:2003fm}. 
The square roots in Eq.~(\ref{eq:edip-polariz1}) are attempts to 
incorporate the pion production threshold behavior, above which 
$\alpha_n$ develops an imaginary part. 
The parameters $\omega_{\Delta}$ and $\Gamma_{\Delta}$, respectively, are quite close to 
the $n$-$\Delta$ mass splitting and the resonance width as 
exhibited in Table~1 of~\cite{BHH17}.
These specific forms also assume the smooth and asymptotically decreasing 
behavior of $\alpha_n$ and $\beta_n$ at 
imaginary frequencies, which are expected from analyticity of the 
Compton $S$-matrix and used in the construction of our Casimir-Polder 
potentials. 
\begin{figure}[tbh]
\begin{center}
\includegraphics[width=0.45\textwidth,clip]{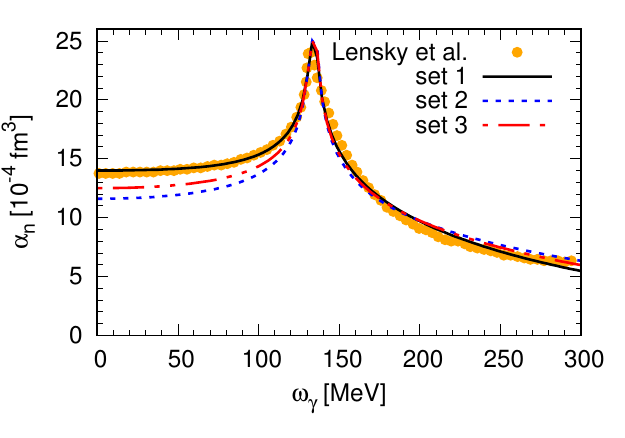}
\includegraphics[width=0.45\textwidth,clip]{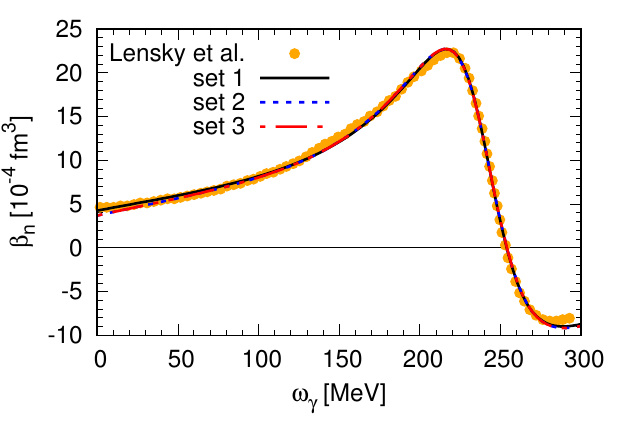}
\caption{\protect 
Dynamic electric (left) and magnetic (right) polarizabilities, as functions 
of the photon energy $\omega_{\gamma}$.
The yellow circles are the CB-$\chi$EFT results of Lensky 
{\em et al.}~\cite{Lensky:2015awa} while sets 1, 2, and 3 correspond to 
our parametrizations using the numbers specified in \cite{BHH17}.}
\label{fig:n-polariz}
\end{center}
\end{figure}

We fit Eqs.~(\ref{eq:edip-polariz1}) and (\ref{eq:mdip-polariz1}) to the 
covariant formulation of baryon chiral effective field 
theory (CB-$\chi$EFT) of Lensky, McGovern, and 
Pascalutsa~\cite{Lensky:2015awa}, which takes proper account of the nucleon 
recoil corrections to all orders. 
For $M_n=938.919$~MeV and letting $M_{\pi}$ be a free parameter we obtain 
$M_{\pi}=134.051$~MeV, which is fairly close to the neutral pion mass 
(134.98 MeV). The remaining 
parameters are given in \cite{BHH17}. 
The parametrizations visibly describe well the results 
of~\cite{Lensky:2015awa} and remain well within the comparatively large 
theoretical uncertainties~\cite{Lensky:2015awa,Hagelstein:2015egb}.
We also checked that our results are in qualitative agreement with chiral 
EFT results at imaginary frequencies up to $iM_{\pi}$~\cite{BHH17}. 

\begin{figure}[tbh]
\begin{center}
\includegraphics[width=0.65\textwidth,clip]{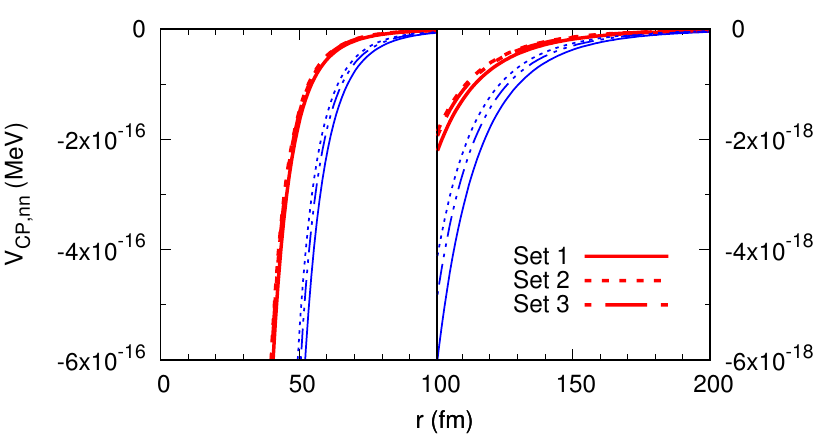}
\caption{\protect 
C-P-interaction for two neutrons, as a function of the separation distance $r$. 
The red thick and blue thin lines correspond to the use of dynamical and 
static dipole polarizabilities, respectively.}

\label{fig:Vcp_nn01}
\end{center}
\end{figure}

From our parametrizations~(\ref{eq:edip-polariz1}), (\ref{eq:mdip-polariz1}) 
we obtain the neutron-neutron C-P-interaction via Eqs.~(\ref{CP-dip-dip}) and 
(\ref{eq:integ_ij}). The results are given in Fig.~\ref{fig:Vcp_nn01} 
as functions of the separation distance. 
The bold red curves correspond to $V_{CP,nn}(r)$ given by the dynamic 
polarizabilities previously shown, while the thin blue curves correspond 
to the static limit $\alpha_n(\omega)$, $\beta_n(\omega)$ $\to$ 
$\alpha_n(0)$, $\beta_n(0)$. 
% Fig. 2
\begin{figure}[tbh]
\begin{center}
\includegraphics[width=0.55\textwidth,clip]{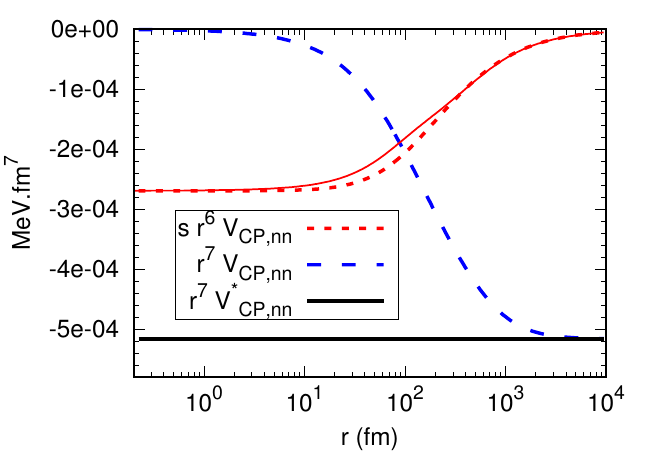}
\caption{\protect 
The neutron-neutron C-P-interaction as a function of the separation distance 
$r$, multiplied by $s\,r^6$ (red dotted line, with $s=100$~fm) and $r^7$ 
(blue long-dashed line). The black solid line is the C-P-potential from 
the static limit of the dipole polarizabilities, multiplied by $r^7$.}
\label{fig:Vcp_nn02}
\end{center}
\end{figure}
It is clear from Fig.~\ref{fig:Vcp_nn02} that the effect of dynamical 
polarizabilities is to decrease the magnitude of the potential up to distances 
as far as 200 fm. The expected long-distance limit of Eq.~(\ref{nn-infty}) 
can be inspected in Fig.~\ref{fig:Vcp_nn02}. 
We use parameters from Set 3, which illustrates 
well the qualitative behavior of the other sets. 
In the red dotted curve   we multiplied the C-P potential by $s\,r^6$, where 
$s=100\,\mathrm{fm}$ to fit in the figure. 
The blue dashed and black solid lines, respectively, are
the dynamic and static polarizabilities versions of $V_{CP,nn}$ (the latter 
indicated by $V^{*}_{CP,nn}$ in the figure), multiplied 
by $r^7$. 
The red thin solid line is the arctan parametrization~\cite{OCaSuc69} 
commonly used in atomic physics (see, for example, Ref.~\cite{FriJacMei02}) 
to make the transition from the $1/r^6$ van der Waals to the asymptotic 
$1/r^7$ Casimir-Polder behavior~\cite{Arn73}. 

The red dotted curve evidences the $1/r^6$ behavior at small distances 
up to $\approx 20$~fm---a region that is dominated by energies larger than 
used to set our parametrizations (\ref{eq:edip-polariz1}), 
(\ref{eq:mdip-polariz1}). This can be checked via the dominance 
of the exponential factor in Eq.~(\ref{eq:integ_ij}): $r\lesssim 20$~fm 
involves photon energies larger than 
$(2\alpha_0\times 20\,{\rm fm})^{-1}\sim 670$~MeV. 
The Delta resonance starts contributing at about 
$(2\alpha_0\omega_{\Delta})^{-1}\sim 50$~fm mostly via $\beta_n(\omega)$, 
which is numerically of $\sim 10$\%. 
Our results can therefore be considered valid for distances beyond 50~fm. 
The same reasoning applies to the contribution of the pion production 
threshold, at around 100~fm. 
The expected asymptotic behavior (\ref{nn-infty}) is only reached for 
$r\gtrsim 10^3$~fm, dominated by dynamic polarizabilities in the region 
$\omega_{\gamma}\lesssim 10$~MeV~\cite{BHH17}. 

In Ref. \cite{BHH17} we have also calculated the neutron-wall C-P 
interaction and the wall-neutron-wall interaction. For completeness we give 
these results below  and further details can be found in \cite{BHH17}. 
\begin{figure}[tbh]
\begin{center}
\includegraphics[width=0.55\textwidth,clip]{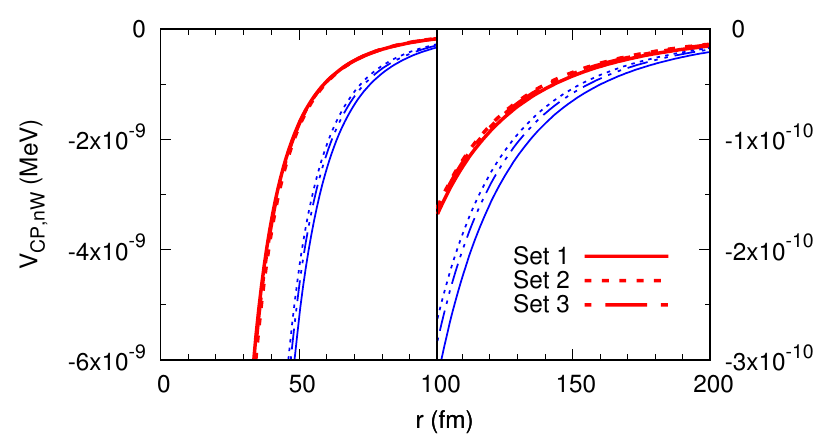}
\includegraphics[width=0.43\textwidth,clip]{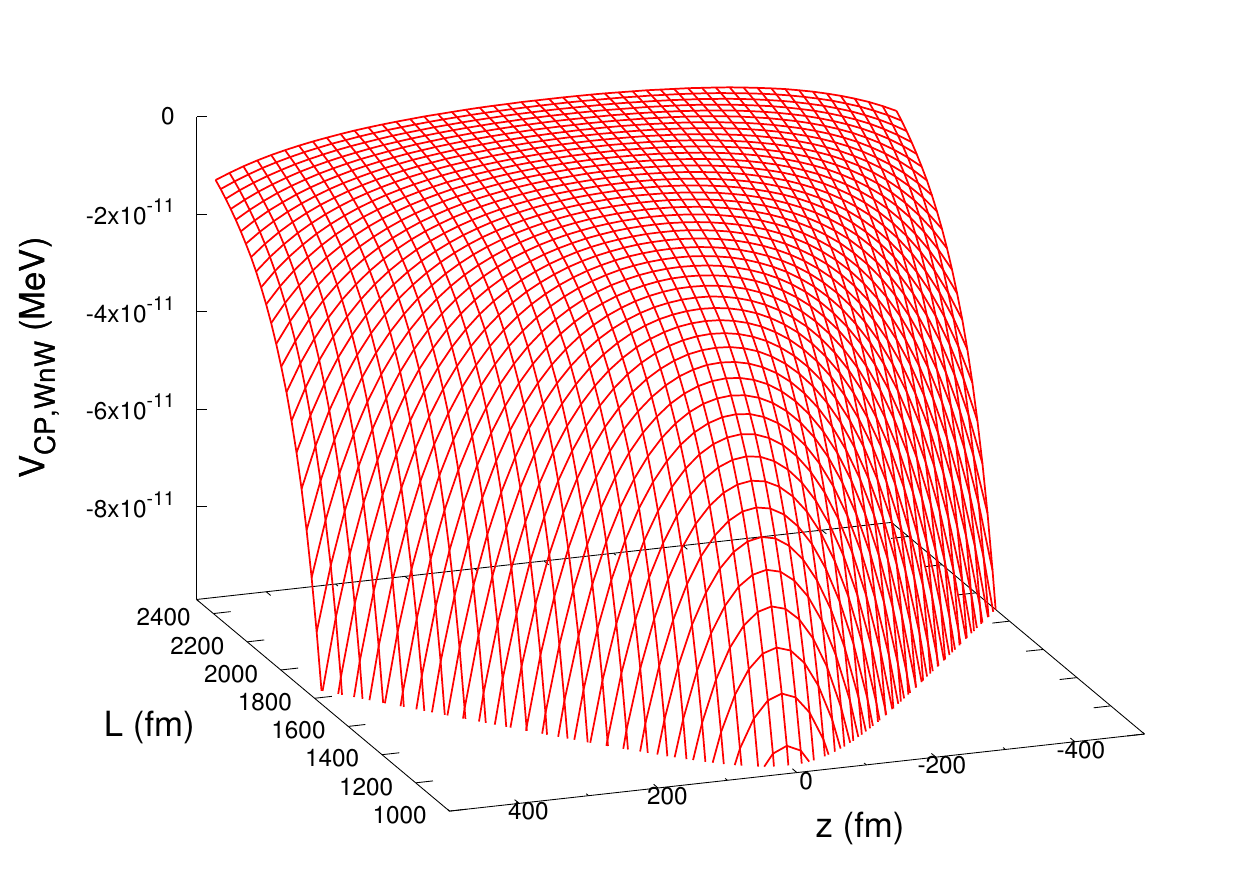}
\caption{\protect
Left panel: C-P-interaction for a neutron and a wall, as a function of the 
separation distance $r$. Notation is the same as Fig.~\ref{fig:Vcp_nn01}.
Right panel: C-P-interaction for a neutron between two walls as a function 
of the neutron position from the midpoint $z$ and the separation between 
the two walls $L$.}
\label{fig:Vcp_nw01}
\end{center}
\end{figure}

\section{The C-P interaction among three and four neutrons}
\label{tetra}
As we have discussed in Section 2, the C-P interaction between two neutrons is not very sensitive to the high frequency dispersive response of the neutron connected with the opening of the one-pion production channel and the excitation of the $3/2^{+} \Delta$ resonance. Nevertheless the overall C-P interaction associated with the usual dipole-dipole dispersive force is appreciable and detectable through careful analysis of low-energy n-n scattering
\subsection{The 3n C-P interaction}
 It is certainly of interest to investigate the C-P interaction among three, four and more neutrons. The usual approach to this few and many-body system through the introduction of a mean field is not appropriate here \cite{Power85}. The non-additive dispersive C-P potentials for three and four neutrons can be read off from this paper, and we give here the final results appropriate for a given geometry. For an equilateral triangular configuration of three neutrons with sides  of length $ r$ the general result of \cite{Power85} would give,
\begin{equation}
V^{CP}_{3n}(r) = \frac{2^{4}\times 79}{3^5}\frac{\hbar c}{\pi}\frac{\alpha_{n}^{3}}{r^{10}}
= 1.73\frac{\hbar c}{\pi}\frac{\alpha_{n}^3}{r^{10}}
\end{equation}
while the linear configuration $n$-$n$-$n$ with the inner $n$-$n$ separation being $r/2$
gives
\begin{equation}
V^{CP}_{3n}(r) = - 186\frac{\hbar c}{\pi}\frac{\alpha_{n}^3}{r^{10}} .
\end{equation}
The prediction of \cite{Power85} for three neutral molecules as extended to neutrons in this paper shows that the geometry plays a central role. The C-P interaction is repulsive in the triangular case while it is attractive in the linear case. This is in contrast to the two-neutron C-P interaction which is universally attractive.

\subsection{The 4n C-P interaction and its potential relevance to the tetraneutron}
We turn now to the C-P interaction in the case of a tetramolecule as derived by \cite{Power85} and as applied here for the 4 neutron system in the spatial configuration of a tetrahedron of edge length $r$,

\begin{equation}
V^{CP}_{4n}(r) = - \frac{3\times41\times2689}{2^{15}}\frac{\hbar c}{\pi}\frac{\alpha_{n}^{4}}{r^{13}}
= - 633\frac{\hbar c}{\pi}\frac{\alpha_{n}^{4}}{r^{13}} ,
\end{equation}
which is universally attractive. 

\begin{figure}[tbh]
\begin{center}
\includegraphics[width=0.55\textwidth,clip]{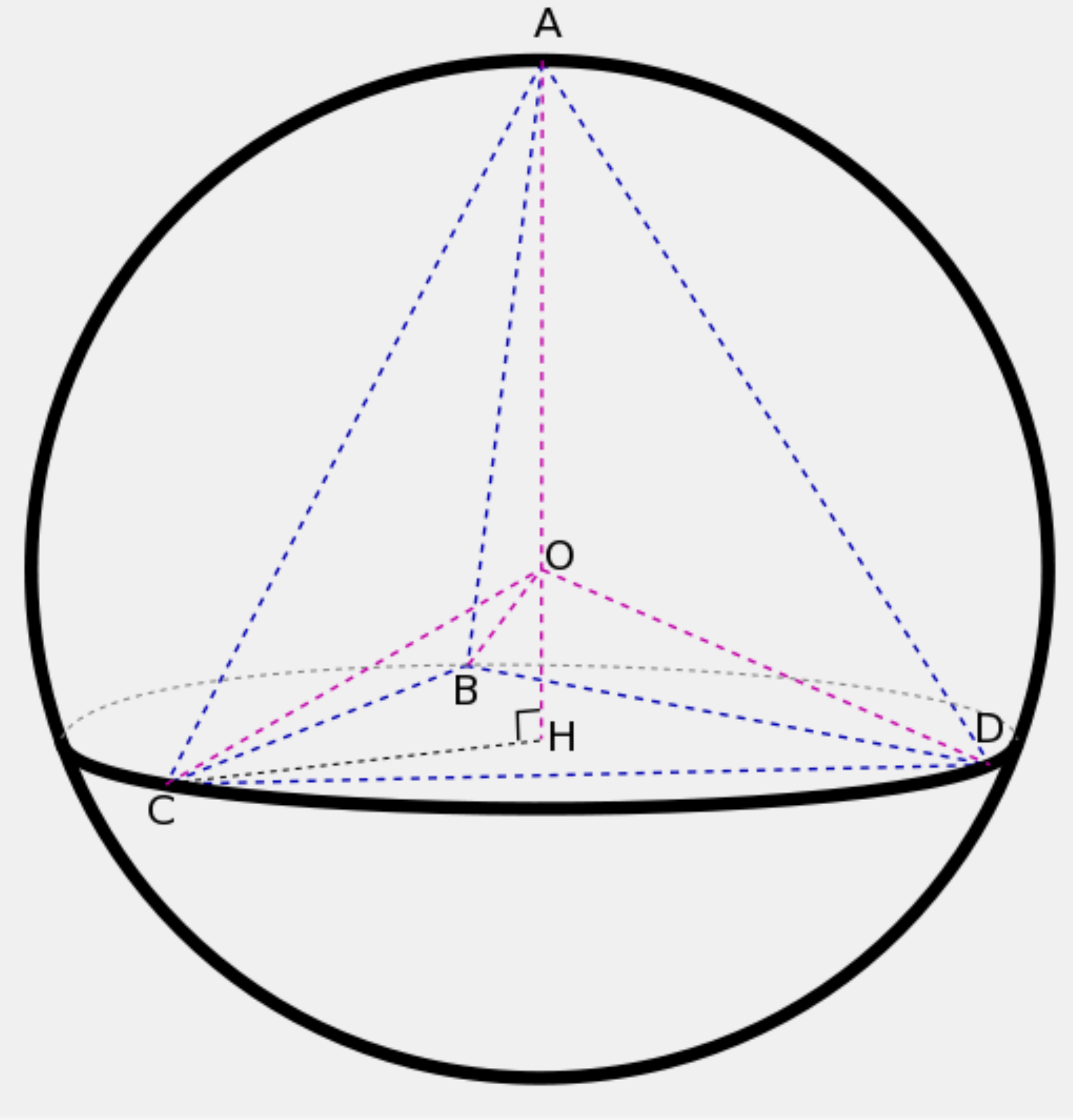}
\caption{\protect 
A tetrahedron with its circumsphere. [Title: Regular tetrahedron and its circumscribed sphere, CC BY-SA 3.0, File: goo.gl/Z6SyYe (unmodified), Author: Illustr, License: https://creativecommons.org/licenses/by-sa/3.0/deed.en.]}
\label{tetra}
\end{center}
\end{figure}

To summarize, with the value of the electric polarizability $\alpha_{n} = 12.6 \times 10^{-4} \mathrm{fm}^3$, we have the following results:

\begin{enumerate}
\item
For the equilateral triangle,
\begin{equation}
V^{CP}_{3n}(r) = 2.12 \times 10^{-7}\frac{1}{r^{10}}     [\mathrm{MeV}]
\end{equation}
\item
For the linear chain, $n$-$n$-$n$ with the $n$-$n$ separation $r/2$,
\begin{equation}
V^{CP}_{3n}(r) = - 2.28 \times 10^{-5}\frac{1}{r^{10}}  [\mathrm{MeV}]
\end{equation}
\item
For the tetrahedron configuration of edge length $r$, we have,

\begin{equation}
V^{CP}_{4n}(r) = - 1.55\times 10^{-9}\frac{1}{r^{13}}       [\mathrm{MeV}]
\end{equation}
\end{enumerate}
In all the equations above $r$ is in femtometer.\\

It is known that there is no bound dineutron. This fact became important in so far as the recent production of Borromean nuclei, such as $^6$He and $^{11}$Li, where the structure is understood as a bound core plus two neutrons. None of the two fragment subsystems  are bound, e.g., $^4$He + $n$ = $^5$He, $n + n$ are unbound, and similarly for $^{11}$Li considered as a stable core, $^9$Li bound to two neutrons; with $^{9}$Li $+ n$ = $^{10}$Li and $n + n = 2n$ being unbound.\\

There are also no bound or observable resonant trineutron states. The cause for this is the Fermi nature of the neutron and the corresponding Pauli exclusion principle.\\

The quest for bound tetraneutrons was started back in the early sixties \cite{ArganI63, ArganII63}, where the possible existence of the bound tetraneutron in the reaction $^4$He$(\gamma$,$\pi^{+}$)$^4$H $\rightarrow$ $^3$H + $n$ was claimed. This result was challenged by \cite{Schiffer63} through the study of the byproducts of the induced and spontaneous fission of uranium and californium. They obtained negative results. 

The search for tetraneutrons was revived with the advent of secondary beams of  very neutron rich unstable nuclei such as $^8$He, $^{11}$Li, $^{14}$Be and $^{15}$B. 
Ref. \cite{Marq2002} studied the reactions of these exotic nuclei with a carbon target, and detected tetraneutrons in the elastic breakup reaction $^{14}$Be + $^{12}$C $\rightarrow$  $^4$n + $^{10}$Be + $^{12}$C. These results posed a great challenge to nuclear structure theory, as Pieper \cite{Pieper03} has pointed out. Pieper used the most realistic nuclear Hamiltonian at the time which predicts successfully many properties of the nucleus, and could not find a bound tetraneutron system. 
 
In 2016, \cite{Kisamori2016}, studied the reaction $^4$He ($^8$He, $^8$Be) and  found a resonant tetraneutron state in the missing mass spectrum. This energy of the tetraneutron resonance was found to be $E_{R} = 0.83 \pm 0.63 (\mathrm{statistical}) \mp 1.25 (\mathrm{systematic})$ MeV above the threshold of four neutron decay. The  width of this resonance, $\Gamma_{R}$ was found to be 2.6 MeV (Full Width at Half Maximum). Clearly, $\Gamma_R > 2E_R$. Three theoretical papers \cite{Hiyama16, Carboneli17, Rimantas17} contested the existence of a tetraneuron resonance as it would require the inclusion in the four body description a strong three-nucleon force which would have a undesirable consequence on the other properties of the nuclear system. On the other hand \cite{Fossez2017} performed a realistic structure calculation which included the coupling to the continuum a resonant $^4$n state at roughly the same energy of \cite{Kisamori2016}, but the width came out to be larger than 3.7~MeV. Since the tetraneutron resonance is a wide resonance in the sense that its width is larger than twice its energy, its decay would deviate appreciably from a simple exponential. We anticipate that the tiny C-P interaction may play a role on the decay and lower the value of the width. To accomplish this a careful calculation of the influence of the C-P force on the basic nn scattering length would be needed and with this a reexamination of the 4n interaction and decay properties can be assessed. This work is in progress.

\section{Conclusions}
In this contribution we have discussed our recent findings about the Casimir-Polder interaction between neutrons. In particular we assessed the importance of the pion production threshold and the $\Delta$ resonance on our dispersive potentials. We also considered the C-P interaction between three and four neutrons. Relevance of our findings about the 
4$n$ C-P interaction on the decay properties of the recently observed tetraneutron resonance is pointed out. Further work on this last point is required to better pin down the role of the $4n$ C-P interaction.

 \section{Acknowledgements}
JFB is supported in part by the U. S. NSF through a grant for the Institute of
Theoretical Atomic, Molecular, and Optical Physics at Harvard University and Smithsonian
Astrophysical Observatory.
RH and MSH are supported in part by the Funda\c c\~ao de Amparo \`a Pesquisa do Estado de
 S\~ao Paulo (FAPESP). MSH is also supported by the
Conselho Nacional de Desenvolvimento Cient\'ifico e Tecnol\'ogico  (CNPq). and by the Coordena\c c\~ao de Aperfei\c coamento de Pessoal de N\'ivel Superior (CAPES), through the CAPES/ITA-PVS program.

\end{document}